\documentstyle[epsfig]{mn}
\begin{document}


\title
[On the fate of gas in ultraluminous infrared galaxies at low and high
redshift]
{On the fate of gas in ultraluminous infrared galaxies at low and high
redshift} 
\author[Neil Trentham] 
{
Neil Trentham$^1$\\
$^1$ Institute of Astronomy, Madingley Road, Cambridge, CB3 0EZ.\\
}
\maketitle 

\begin{abstract} 
It is often suggested that the 
distant galaxies recently identified
in 850-$\mu$m surveys with the SCUBA bolometer array on the JCMT 
Telescope are high-redshift analogues to local ultraluminous
infrared galaxies, based on their similar spectral energy
distributions and luminosities.
We show that these two populations of objects must
differ in at least one fundamental way from each other. 
This assertion is based on a consideration of the
possible fates of gas in the high redshift SCUBA galaxies, given the
requirement that they most evolve into some subset of the low-redshift
galaxy population with a comoving density of about 10$^{-4}$ Mpc$^{-3}$. 
One possibility is that the SCUBA galaxies have similar
gas density profiles to local ultraluminous galaxies.  If this is the
case, then they must derive almost all their power from AGNs, which appears
not to be the case for local ultraluminous galaxies, which are predominantly
star-formation powered.  Another possibility is that the SCUBA galaxies
have more extended gas density profiles than local ultraluminous galaxies.
In this case they must be almost all star-formation powered, and much of
the star formation in the Universe can happen in these objects.  Either way
there is a significant difference between the low- and high- redshift
populations.  

\end{abstract} 


\section{Ultraluminous galaxies at low redshift}  

Ultraluminous infrared galaxies (ULIGs) are peculiar galaxies with infrared 
luminosities in excess of 10$^{12}$ L$_{\odot}$
(Sanders \& Mirabel 1996).  These galaxies have
very dense gas cores, as indicated by molecular line measurements,
both of CO (Sanders, Scoville \& Soifer 1991), and also of high-density
traces like HCN (Solomon, Downes \& Radford 1992) and CS (Solomon,
Radford \& Downes 1990).  A number of measurements at radio, infrared,
and optical wavelengths 
(Condon et al.~1991, Surace et al.~1998,
Surace \& Sanders 1999, Soifer et al.~1999)   
suggest that most of the very substantial bolometric luminosities
of these galaxies comes from these dense molecular cores. 

The existence of these extreme objects leads to the question: what do they
evolve into?  One approach to answering this question is to look for
stellar systems at redshift
$z=0$ with central densities (see Fig.~1 of Binggeli 1994
for measrements of stellar densities in local galaxies) 
of about 100 M$_{\odot}$
pc$^{-3}$ or greater, the gas densities of the ULIG cores. 
Kormendy \& Sanders (1992) made this comparison and showed that the only
realistic candidates on these grounds at $z=0$ are giant elliptical 
(gE) cores. 
Galaxy disks (which have lower surface-brightnesses
than gE cores; Freeman 1970) do not
have high enough stellar densities.  
These gE cores are dense but
have profiles that are flatter than an extrapolation of
a de-Vaucouleurs $r^{{1}\over{4}}$ law to small radii; they are rarely
isothermal (Lauer 1985).  Recent $Hubble$ $Space$ $Telescope$ imaging
(Faber et al.~1997) has confirmed the existence of such cores in the
most luminous ellipticals (they are, however, less common in lower-luminosity
ellipticals).  
Often the cores are kinematically decoupled from the rest of the galaxy
(e.g.~Forbes et al.~1996), 
suggesting that their formation happens in a different
way to that of the rest of the galaxy.  These cores also tend to have
supermassive black holes (BHs) in their centers, 
as inferred from stellar-kinematical measurements.  The
bigger the galaxy, the bigger this BH
(Magorrian et al.~1998, van der Marel 1999, Ferrarese \& Merritt 2000,
Gerbhardt et al.~2000). 
Additionally, the bigger the
galaxy, the bigger the core (Lauer 1985).
 
An additional connection between the dense molecular cores of the
ULIGs and the gE stellar cores + BHs is that their
masses and sizes are similar, in addition to their densities.
Typical masses for both are several times 10$^9$ M$_{\odot}$ and typical
radii are several hundred parsecs (see Sakomoto et al.~1999 for details of
Arp 220, the nearest ULIG, and Lauer et al.~1985 for the core properties of
gE cores).  The BH contribution to the gE core masses
can be added in from the correlations of
Magorrian et al.~1998 or van
der Marel 1999 which relate the BH mass to the galaxy mass, in combination
with Tables 2 and 3 of Lauer 1985, which relate the core size to the size
of the whole galaxy.  Masses of the stellar cores can be computed
directly from the core sizes, along with the mass-to-light ratio of
Fukugita, Hogan \& Peebles (1998).

But locally, gE galaxies are much more common than ULIGs.  The number
density of ULIGs with far-infrared luminosities in excess of 10$^{12}$ 
L$_{\odot}$ is about $1.1 \times 10^{-7} {\rm Mpc}^{-3}$
(Saunders et al.~1990).  Typical ULIGs
have total
$K$-band 
(2.2 $\mu$m) absolute
magnitudes of $M_K = -25.4$ (Carico et al.~1988, 1990;
here, as throughout this paper we assume an Einstein-de Sitter
cosmology with $H_0$ = 50 km s$^{-1}$ Mpc$^{-1}$),
most of which comes from pre-existing stars in the progenitor galaxies
(the dense cores are optically thick at $K$-band; Goldader et al.~1995).
The number density of normal galaxies 
with $M_K < -25.4$ is
$1.6 \times 10^{-4} {\rm Mpc}^{-3}$ (Szokoly et al.~1998), a factor
of about 1500 larger than the density of ULIGs.  Most of these 
$K$-band-luminous normal galaxies are early-type galaxies, which have the
cores + BHs
discussed in the previous two paragraphs; the contribution of
luminous disks, which tend to be blue,
to the $K$-band luminosity function at the very bright end is small (see
e.g.~Binggeli, Sandage \&
Tammann 1988 for the optical luminosity function of the
different Hubble types, and Huang et al.~1997 for the optical-near-infrared
colors).  

Therefore if gE cores were made by ULIGs, the comoving density of ULIGs
must have been a factor 1500 higher in the past than it is today,
approximately 10$^{-4}$ Mpc$^{-3}$.
But this is 
very similar to the comoving density (e.g.~Trentham, Blain \& Goldader
1999) of the SCUBA ($>1$ mJy at
850 $\mu$m in the observer frame) sources 
(Smail, Ivison \& Blain 1997, Barger et al.~1998, Hughes et al.~1998,
Eales et al.~1999), 
which are presumably
high-redshift objects, and which Barger et al.~(1998) argue are 
ULIGs based on their similar bolometric luminosities and spectral
energy distribution (SEDs). 
Therefore a picture in which SCUBA sources make gE cores and their associated
black holes seems very likely not just based on similarities of their SEDs
and luminosities to those of local
ULIGs but also due to the comoving number density of local gE cores
+ BHs being
similar to that of the high-redshift SCUBA galaxies. 
It is this possibility that we investigate in the rest of this
paper.

\section{Ultraluminous galaxies at high redshift}

In the previous section we argued that the SCUBA sources are probably
the high-redshift analogues of low-redshift ULIGs, both in terms of
their SED and luminosity concordance, and also in terms of the comoving number
density concordance between the SCUBA galaxies and the likeliest remnants
of the local ULIGs (the gE cores + BHs).

This SED (Ivison et al.~1998) and luminosity (Barger et al.~1999) concordance
is observed for individual sources, and is also implied on statistical
grounds.   
By this it is meant that: if the SCUBA sources were significantly  
hotter than local ULIGs, 
then they would overproduce the far-infrared background shortward of about 500
$\mu$m measured by Fixsen et al.~(1998) using $COBE$.  
If they were signifiantly colder than local ULIGs, they would severely
underpredict the background at 450 $\mu$m:
the SCUBA sources are observed to
generate at least one-third of the background measured by Fixsen et
al.~at 450 $\mu$m
(Blain et al.~1999a).  These two statements assume a median source
redshift of approximately 3 (see the radio measurements of Smail
et al.~2000) and then follow directly from the result that
almost the entire background
at 850 $\mu$m is generated by the SCUBA sources (Blain et al.~1999b,
Barger, Cowie \& Sanders 1999).  Both statements 
are independent of any specific model of
SCUBA source luminosity and density evolution.   

Now let us make the following additional assumption, which we will
call {\bf A1}: 
\vskip 3pt
\noindent
{\bf The density structure of the gas in the SCUBA sources is the same as in
local ULIGs.} 
\vskip 3pt
This is an assumption, not an observation,
since we have no direct probes of the gas densities in
the SCUBA sources.  We only know about CO in two of the SCUBA sorces (Frayer 
et al.~1998, 1999), and have no measurements at all of high-density tracers
like HCN or CS in these objects. 
There are two indirect pieces of evidence supporting {\bf A1}.
Firstly, HCN has been detected in the 
Cloverleaf quasar (Barvainis et al.~1997), a strongly-lensed infrared quasar
which has an SED at far-infrared wavelengths similar to those of the
SCUBA sources.
Secondly, if the burst producing the high far-infrared luminosity is shortlived
(see e.g.~Blain et al.~1999a), 
then it is unlikely that it can be maintained
in a coherent way over a galactic scale (in this case, the large scale
over which the luminosity emerges from requires that the ultimate power
source is star formation).  Therefore the gas fuelling the
burst is unlikely to be distributed over the whole galaxy. 
Scenarios do exist in which 
positive feedback can happen where explosions in one part of a
galaxy can trigger those in another part (e.g.~Taniguchi,  
Trentham \& Shioya
1998), but these mechanisms are not powerful enough to generate the kinds
of luminosity required here over a whole galaxy. 

The important point is that if
{\bf A1} is correct, then the above suspicion of the connection between local 
ULIGs and the SCUBA sources based on the 
(remnant) number density and SED concordance
now becomes a very strong assertion.  This is because the {\it only
things that the SCUBA sources can evolve into are elliptical galaxy
cores and their associated black holes}. 

We do not
consider the possibility that stars are made in the nucleus of the galaxies (as
would be required by {\bf A1}) and then redistributed throughout the 
galaxy in stellar-kinematical mergers 
because (1)
the dense stellar cores so produced will not be disrupted by mergers (or
secular evolution), (2) we would require about 100 mergers per
L$^{*}$ elliptical and the observed merger rate 
(e.g.~Carlberg et al.~2000) is not that high, 
and (3) the number density concordance described
above is then lost due to the
large number of mergers.

Van der Marel 1999 finds
the black hole mass (in solar units) $\log_{10} M_{\rm BH} \approx -1.83 +
\log_{10} L_V$.  Therefore for a Schechter (1976) $L^{*}$ elliptical
galaxy ($M_V = -21.5$) at redshift $z=0$, the central black hole has mass
$5.1 \times 10^{8} {\rm M}_{\odot}$.  For the same elliptical galaxy,
the stellar core mass is $M_{*c} = 1.6 \times 10^{9} {\rm M}_{\odot}$, 
assuming that the
core is approximately isothermal at small radii ($r<r_c$ where $r_c$
is the core radius), that the galaxy has
a de Vaucouleurs $r^{1/4}$ profile at large
radii, that the ratio of $r_c$ to the de Vaucouleurs effective
radius $r_e$ is 0.033 (derived from the sample of Lauer 1985), and
a value of $\left(M/L_V \right) = 3.86$ for elliptical galaxies
(see the discussion in Section 2.1.3 of Fukugita et al.~1998 -- this
assumes the stellar 
initial mass function (IMF) of Gould, Bahcall \& Flynn 1996, which turns over
at about 0.5 M$_{\odot}$, similar to the IMF of Kroupa, Tout \& Gilmore 
(1993) that we adopt below).   
Therefore, for typical luminous elliptical galaxies 
\begin{equation} 
{{M_{\rm BH}} \over{ M_{*c}}} \approx {{1}\over{3}} 
\end{equation} 
Given the scatter in the correlations presented by van der Marel and
Lauer, and the fact that many cores deviate significantly from
isothermality, 
the most extreme galaxies may deviate from this ratio by as much
an order of magnitude.  

For accretion onto a black hole, 
\begin{equation}
L_{\rm acc} = \eta \dot{M} c^2
\end{equation}
where $\eta$ is the efficiency of accretion and $\dot{M}$ is the mass
accretion rate.

For dust-enshrouded star formation where almost all the flux is absorbed
and reradiated at far-infrared wavelengths, the star formation rate (in
M$_{\odot}$ yr$^{-1}$) is 
(Rowan-Robinson et al.~1997)  
\begin{equation}
\dot{ M_{*}} = 1.2 \times 10^{-10} L_{*}. 
\end{equation}
where $L_{*}$ (in L$_{\odot}$)
is the total luminosity generated by the young stars.
This relation adopts a value of $\phi = 0.45$ appropriate to the stellar
IMF of Kroupa et al.~(1993), 
and a value of $\epsilon = 1$ (see 
Section 3 of
Rowan-Robinson et al.~1997 for a definition of those parameters and
a discussion of the derivation of this equation and other approaches). 
The IMF chosen here has been shown to be consistent with most current
observations of both star-forming regions and the solar
neighbourhood (Gilmore \& Howell 1998). 

Combining the two previous equations yields
\begin{equation}
L_{\rm acc} / L_{*} \sim 1700 \, \eta .
\end{equation}
Accretion is a lot more efficient than star-formation at generating
luminosity per unit mass: a 10$^{12}$
L$_{\odot}$ galaxy has a star formation rate
 of about 80 M$_{\odot}$ yr$^{-1}$ but a
2 $\times$ 10$^{12}$ L$_{\odot}$ quasar has an accretion rate  
of about 1 M$_{\odot}$ yr$^{-1}$ for $\eta = 0.1$ (e.g.~Chokshi
\& Turner 1992).  
Therefore averaged over time in the SCUBA sources:
\begin{equation}
< L_{\rm acc} / L_{*} >_{t} \sim \, 560 \eta
\end{equation}
This ratio is very big, meaning that the total power generated 
in the SCUBA sources comes almost
entirely from accretion if {\bf A1} is correct. 
Note that throughout this calculation, we have been working backwards,
deriving the properties of the progenitor from the properties of the
remnants under the assumption that we can match particular progenitors
to particular remnants; 
this is quite a different calculation from one in which we
start out with a cloud of gas and attempt to track its progress.

The fact that the power sources are so heavily accretion-dominated means that
most SCUBA sources must be dust-enshrouded active galactic nuclei (AGNs) if
{\bf A1} is correct.  This is not surprising, since the SCUBA sources
contribute at least 10 per cent of the bolometric luminosity density of the
Universe, and this could not be generated from the stars in gE cores alone,
which comprise less than 1 per cent by mass of the stars in the Universe,
and these are the only stars that the SCUBA sources are allowed to make
if {\bf A1} is correct.

\section{Additional evidence for the SCUBA source - AGN connection}

There exist other reasons for suggesting a connection between the
SCUBA sources and dust-enshrouded AGNs.  

\vskip 4pt
\noindent
{1.~}The local supermassive BH density (Magorrian et al.~1998,
van der Marel 1999) 
is high and its production generates 
a bolometric background of 
$$5.1 \,\left( {{\rho_{\rm AGN,dusty}} \over {2 \times 10^5 \, 
   {\rm M}_{\odot} {\rm Mpc}^{-3}
   } }\right)
  {\rm nW}{\rm m}^{-2}{\rm sr}^{-1}$$
(Trentham \& Blain, 2000) if they radiate at one-tenth of
the Eddington luminosity and   
if the luminosity density of obscured AGNs follows that
measured by Boyle \& Terlevich (1998) for optical quasars.
This equation follows from a consideration of the energy released by
accretion processes per comoving volume element given a final MDO
density some part $\rho_{\rm AGN,dusty}$ of which was generated by 
dust-enshrouded accretion.  
The fiducial ${\rho_{\rm AGN,dusty}}$ is the estimate of Salucci et al.~(1999)
for obscured quasars.
The SCUBA sources in Fig.~1 generate about 
7 ${\rm nW}{\rm m}^{-2}{\rm sr}^{-1}$, which is close to this value. 

\vskip 4pt
\noindent
{2.~}There is growing evidence that the
hard (30 keV) X-ray background originates from absorbed
quasars which reradiate energy
absorbed at optical, ultraviolet, and soft X-ray at far-infrared
and submillimetre wavelengths.  Were this the case, Fabian \& Iwasawa
(1999) find a total reradiated energy density of about
$ 3 \, {\rm nW}{\rm m}^{-2}{\rm sr}^{-1}$.  This is close to the
numbers in the previous paragraph, and it is therefore plausible that
these same AGNs that we are hypothesizing are the SCUBA sources are also
the objects which contribute to the hard X-ray background (see also
Almaini et al.~1999 and Bautz et al.~2000).  Note that if Compton-thick
sources are very common (see Maiolino et al.~1998), the reradiated
energy will be somewhat higher than $ 3 \, {\rm nW}{\rm m}^{-2}{\rm sr}^{-1}$.

\vskip 4pt
\noindent
{3.~}The Madau plot -- that is, the cosmic star-formation rate of the Universe
as a function of redshift -- when determined by ultraviolet and
optical measurements alone (e.g.~Steidel et al.~1999) and integrated
over time, produces a total stellar density in critical units of  
$\Omega_* \sim 0.004$ (Pettini 1999), equal to the observed local value
(Fukugita et al.~1998)
Therefore an additional population of objects which are star-forming would
lead to $\Omega_*$ being overproduced.  Since $> 10$ per cent of the
bolometric luminosity of the Universe is generated in the SCUBA sources,
were they to be star-forming, they would be producing a substantial
fraction of $\Omega_*$.  We therefore require an alternative power
source, such as accretion onto a black hole.  
The Madau plot described above includes a correction for star-formation
obscured by dust in normal galaxies, which are not the same as the
ULIGs or SCUBA sources described in the previous sections: their predicted
850-$\mu$m fluxes, even for the most luminous normal star-forming
galaxies, are substantially lower (see e.g.~Trentham \& Blain 2000).

\vskip 4pt
\noindent
{4.~}All 
supermassive BHs at low $z$ are found in spheroids 
(Magorrian et al.~1998), 
except those
in ULIGs like Mrk 231 which are probably in the process of
forming spheroids (the
core in the dissipative collapse, the rest of the galaxy in the 
stellar-kinematical merger of pre-existing
stars).  No supermassive BHs are known in pure disks.
Such a scenario follows from the physical processes described in Section 2.

\vskip 4pt
\noindent
{5.~}In their spectroscopic survey, Barger et al.~(1999) find some 
dust-enshrouded AGNs.
But it is clear that powerful Seyfert 1 galaxies are not present in
substantial number, so that if this sample is AGN dominated, they must be very
highly obscured.  Such objects are not very common in local ULIG
samples (see Section 3.2 of Lutz, Veilleux \& Genzel 1999).

\section{Problems with the SCUBA source - AGN connection}

The previous two sections have provided 
some evidence in support of a picture where {\bf A1} is true.
But there are problems with such a scenario:

\noindent
\vskip 3pt
\noindent
{1.~}The ratio of the number of
ULIGs powered by accretion onto a BH to the number
of ULIGs powered by star formation must 
change from about 1/3 at $z=0$ (Genzel et al.~1998)
to about 80 at the redshift of the SCUBA
sources (for $\eta = 0.1$) if {\bf A1} is correct. 
This is very large effect, and there exists no clear physical mechanism
to account for such a big difference, particularly since the  
low-redshift ULIGs and high-redshift SCUBA galaxies are 
similar in every other respect if {\bf A1} is true.
 
\vskip 3pt
\noindent
{2.~}The $q$ value (relating the far-infrared flux to the 4.85 GHz radio
flux; Condon, Frayer \& Broderick 1991)
of SMM J02399$-$0136 is high, suggesting that
about half of its power (at most 3/4) comes from star formation (Frayer
et al.~1998).
Recent X-ray measurements of this galaxy with {\it Chandra}     
(Bautz et al.~2000) also suggest that the fraction of the bolometric
luminosity contributed by star-formation is at least 20 \%, probably
more.  This is much higher than the value of $1/(560 \eta + 1)$ predicted in 
Section 2. 
For this not to be a concern, SMM J02399$-$0136 must be a
very unrepresentative
example of a SCUBA galaxy
(which is not the case given the optical spectroscopy of Ivison et al.~1998
and Barger et al.~1999) or
its AGN must emit an amount of radio luminosity given its bolometric
luminosity that somehow mimics starbursts.

\vskip 3pt
\noindent
{3.~}If the SCUBA sources are powered by AGNs,
the infalling gas must lose at least
99 \%  of its initial angular momentum before reaching the 
BH event horizon.  This is not achiveable by any known mechanism. 
Suggested mechanisms include global gravitational or magnetic processes 
(Begelman 1994), but the details are unclear. 
Note that the radius of the orbit of the two 
nuclear gas disks in Arp 220 is about
250 pc (Sakamoto et al.~1999), which is approaching the size scale on
which the central black hole dominates the gravitational potential.
Although this has been a long-standing problem, the increased angular
resolution of the gas achieved by current inteferometry gives it an
additional importance if {\bf A1} is to
be regarded as a plausible hypothesis. 

One possible solution that simultaneously addresses all three
problems
above could be that: the ULIGs undergo an intense star formation
phase in the center that generates a bolometric luminosity of 
$10^{12} {\rm L}_{\odot}$ followed 
immediately by an AGN phase that lasts for
a shorter time, but generates a higher bolometric luminosity, say
about $10^{13} {\rm L}_{\odot}$.  If one selects objects 
instantaneously with
bolometric luminosities above $10^{12} {\rm L}_{\odot}$, one would
end up with a sample dominated by starbursts, perhaps in a ratio
similar to what Genzel et al.~(1998) found locally if the relative
timescales and luminosities between the starburst and AGN phase
are extremely large and small, respectively.  
This could remain true even if the time-averaged luminosity was
generated predominantly by AGNs, as required by
eq.~(5). 
Such a scenario could help the angular momentum problem since the
black hole could accrete the stars themselves (Begelman 1994);
given eq.~(1) about one-fourth by mass of the total stars produced would
have to be accreted.  It is also consistent with a picture in which
the most luminous things that exist at any redshift are AGN-powered
(see e.g.~Sanders \& Mirabel 1996).

Another possible solution is that the fraction of ULIGs at $z=0$ that are
AGNs has been underestimated because the AGNs are so heavily
obscured that they are invisible even at mid-infrared wavelengths.  
The detection of hard X-ray emission
from NGC 6240 (Vignati et al.~1999) suggests that the bolometric luminosity
of this infrared-luminous galaxy, previously thought to be 
starburst-powered, originates from an AGN.  If this  
object is at all typical of ULIGs, then the possibility of a much higher
ratio of local ULIGs that are AGN-powered than one-third seems possible
(see however the discussion in the final section of Lutz et al.~1999).
 
\section{Discussion and summary}

We have shown that 
if the SCUBA sources, which appear on SED and luminosity grounds to be
ULIGs at high $z$ have the same gas density structure as local ULIGs,
then they must generate most of their power from embedded AGNs.  Such
a scenario is attractive for many reasons: it generates the correct local
number density of elliptical galaxy cores of the
correct mass, the correct local mass density
of their associated black holes, and gives a formation mechanism for
kinematically decoupled cores
(which probably form in some other way from the rest of the galaxy).
These SCUBA sources would then be very different from local ULIGs,
which derive most of their power from star formation.

But there are problems (see Section 4).  If these cannot be resolved, then
{\bf A1} must be abandoned.  In this event, the gas in the
SCUBA sources must be far more diffuse than in local ULIGs because in
order to generate such huge luminosities, a substantial fraction of
the stars in the galaxies must be made; 
in $z=0$ galaxies, large
fractions of the stars in galaxies do not
reside at extremely high densities of 100 M$_{\odot}$ pc$^{-3}$ 
or greater so it
is improbable that they were made out of gas at these densities.  
In local ULIGs all the bolometric luminosity seems to come out of regions
associated with gas at this high density.

Either way, there is a big difference between local ULIGs and high-redshift
SCUBA sources.  This is the main result of this letter.

Testing the validity of {\bf A1} will be an important exercise in the
future.  HCN measurements (like for
the Cloverleaf quasar) would be best, but until
these and other millimetre wave spectral diagonstics (other than CO) are
available, the best hope is to rely on indirect methods like those outlined
in the letter.

\end{document}